%% file: main.tex
\documentclass[fleqn,usenatbib]{mnras}

\usepackage{newtxtext,newtxmath}

\usepackage[T1]{fontenc}
\usepackage{ae,aecompl}

\usepackage{graphicx}	
\usepackage{amsmath}	
\usepackage{amssymb}	

\usepackage{color, colortbl}
\definecolor{Gray}{gray}{0.9}

\usepackage{afterpage}

\usepackage[switch]{lineno}
\newcommand*\patchAmsMathEnvironmentForLineno[1]{%
  \expandafter\let\csname old#1\expandafter\endcsname\csname #1\endcsname
  \expandafter\let\csname oldend#1\expandafter\endcsname\csname end#1\endcsname
  \renewenvironment{#1}%
     {\linenomath\csname old#1\endcsname}%
     {\csname oldend#1\endcsname\endlinenomath}}%
\newcommand*\patchBothAmsMathEnvironmentsForLineno[1]{%
  \patchAmsMathEnvironmentForLineno{#1}%
  \patchAmsMathEnvironmentForLineno{#1*}}%
\AtBeginDocument{%
\patchBothAmsMathEnvironmentsForLineno{equation}%
\patchBothAmsMathEnvironmentsForLineno{align}%
\patchBothAmsMathEnvironmentsForLineno{flalign}%
\patchBothAmsMathEnvironmentsForLineno{alignat}%
\patchBothAmsMathEnvironmentsForLineno{gather}%
\patchBothAmsMathEnvironmentsForLineno{multline}%
}

\newcommand{\avg}[1]{\langle #1 \rangle}

\newcommand{\Rmin}{R_{\mathrm{min}}}
\newcommand{\Rmax}{R_{\mathrm{max}}}
\newcommand{\zmin}{z_{\mathrm{min}}}
\newcommand{\zmax}{z_{\mathrm{max}}}

\newcommand{\Rr}{\mathrm{R^{r}}}
\newcommand{\Du}{\mathrm{D^{u}}}
\newcommand{\Dr}{\mathrm{D^{r}}}
\newcommand{\DD}{\Du \Dr}

\newcommand{\DRu}{\Du \Rr}

\newcommand{\Pz}{Photo-$z$}
\newcommand{\pz}{photo-$z$}
\newcommand{\pzs}{photo-$z$'s}

\newcommand{\redmagic}{redMaGiC}
\newcommand{\metacal}{\texttt{METACAL}}
\newcommand{\imshape}{\texttt{IM3SHAPE}}
\newcommand{\mof}{\texttt{MOF}}
\newcommand{\bpz}{\texttt{BPZ}}
\newcommand{\dnf}{\texttt{DNF}}

\newcommand{\fiducial}{DES Y1}

\newcommand{\nz}{n^{i}_{\mathrm{PZ}}}

\newcommand{\wz}{clustering-$z$}
\newcommand{\Wz}{Clustering-$z$}
\newcommand{\WZ}{Clustering-$z$}

\newcommand{\spec}{\texttt{COSMOS}}
\newcommand{\Spec}{\texttt{COSMOS}}
\newcommand{\SPEC}{\texttt{COSMOS}}

\newcommand{\combined}{combined}

\defcitealias{xcorrtechnique}{G17}
\newcommand{\gatti}{\citetalias{xcorrtechnique}}

\usepackage{eso-pic}

\AddToShipoutPictureBG*{%
  \AtPageUpperLeft{%
    \hspace{0.75\paperwidth}%
    \raisebox{-3.5\baselineskip}{%
      \makebox[0pt][l]{\textnormal{DES 2017-0263}}
}}}%

\AddToShipoutPictureBG*{%
  \AtPageUpperLeft{%
    \hspace{0.75\paperwidth}%
    \raisebox{-4.5\baselineskip}{%
      \makebox[0pt][l]{\textnormal{FERMILAB-PUB-17-384-AE}}
}}}%

\makeatletter
\def\blfootnote{\xdef\@thefnmark{}\@footnotetext}
\makeatother

\title[DES Y1: WZ Calibration]{Dark Energy Survey Year 1 Results:  Cross-Correlation Redshifts in the DES -- Calibration of the Weak Lensing Source Redshift Distributions}

\input{authors.tex}

\date{Accepted XXX. Received YYY; in original form ZZZ}

\pubyear{2017}

\begin{document}
\label{firstpage}
\pagerange{\pageref{firstpage}--\pageref{lastpage}}
\maketitle

\begin{abstract}
    We present the calibration of the Dark Energy Survey Year 1 (DES Y1) weak lensing source galaxy redshift distributions from clustering measurements.
    By cross-correlating the positions of source galaxies with luminous red galaxies selected by the \redmagic\ algorithm we measure the redshift distributions of the source galaxies as placed into different tomographic bins.
    These measurements constrain any such shifts to an accuracy of $\sim0.02$ and can be computed even when the clustering measurements do not span the full redshift range.
    The highest-redshift source bin is not constrained by the clustering measurements because of the minimal redshift overlap with the \redmagic\ galaxies.
    We compare our constraints with those obtained from \texttt{COSMOS} 30-band photometry and find that our two very different methods produce consistent constraints.
\end{abstract}

\blfootnote{Affiliations are listed at the end of the paper.}
\setcounter{footnote}{1}



\section{Introduction}

The Dark Energy Survey (DES) Year 1 (Y1) Key Project \citep{keypaper} constrains cosmological parameters by combining three distinct measurements of the growth of large scale structure over 1321~deg$^2$:
first, measurements of the weak lensing shear fields \citep{shearcorr} from the cross-correlations of the measured shapes of 26~million ``source'' galaxies divided into four redshift bins \citep{shearcat} from $0.2 < z < 1.3$;
second, cross-correlations of source galaxy shapes \citep{gglpaper} with the positions of 650,000 luminous red (``lens'') galaxies at $0.15 < z < 0.9$ as determined by the \redmagic\ algorithm \citep{redmagicSV};
third, the auto-correlations of the positions of \redmagic\ galaxies \citep{wthetapaper}.
Undergirding our cosmological constraints are estimates of the redshifts of these galaxies from DES photometry in the $griz$ bands. These photometric redshifts (``\pz'') are used twice: first to assign galaxies to tomographic bins, and then to determine the normalised redshift distribution $n^i(z)$\ of galaxies in the $i$-th bin.

This paper describes the calibration of the source galaxy redshift distributions used in the DES Y1 Key Project by looking at their cross-correlations with high-fidelity lens galaxy photometric redshifts. In analogy with photometric redshifts, we shall refer to these estimates of the redshift distribution as ``\wz''.
The redshift distributions are crucial for the prediction of the observable cosmological signals, and their uncertainties must be propagated into our cosmological parameter inference pipeline. For our measurement precision, the most important parameter of a source galaxy redshift distribution is its mean redshift. We focus here on the calibration of that parameter using angular cross-correlations with \redmagic\ galaxies.
\citet[henceforth \gatti]{xcorrtechnique}\ describe how we use simulations to estimate the systematic uncertainties in this method.
\citet{photoz} describe the binning and redshift determination of source galaxies as well as their validation with 30-band \texttt{COSMOS} redshifts \citep{Laigle}.
The assignment and validation of lens galaxy redshifts are described in
\citet{redmagicSV}, \citet{wthetapaper}, and \citet{redmagicpz}.

It has now been almost a decade since \citet{Newman:2008aa} first demonstrated on simulations the use of angular cross-correlations with finely-binned, high-fidelity-redshift galaxies to determine redshift distributions, and over a decade since \citet{Schneider:2006aa} proposed using galaxy angular two-point correlation functions to determine redshift distributions. Since then, the method has been applied to both simulation and real data \citep{Menard:2013aa, Schmidt:2013ac, McQuinn:2013aa, Rahman:2015aa}. Of particular relevance is the recent work in \citet{Hildebrandt:2017aa}, \citet{Johnson:2017aa}, and \citet{Morrison:2017aa}, where \wz\ methods were applied to Kilo-Degree Survey photometric data by cross-correlating with spectroscopic redshifts from the Galaxy And Mass Assembly (GAMA) survey and the Sloan Digital Sky Survey (SDSS). These papers used \wz\ distributions as alternative redshift distributions to those from photometric techniques, and demonstrated the viability -- and the potential -- of using \wz\ methods to determine redshift distributions.
In our present work, we make two significant modifications. First, instead of spectroscopic redshifts over a minimal area in the sky (often only 10-100 deg$^2$), we use the high-fidelity photometric redshifts determined by the \redmagic\ algorithm to measure \wz\ over our entire 1321 deg$^2$ footprint. This is also different from our previous work on DES Science Verification data in \citet{Davis:2017aa}, where we instead used redMaPPer galaxy clusters \citep{Rykoff:2014aa,Rykoff:2016aa}. \redmagic\ redshifts are more than sufficiently accurate for our purposes, and we have many more \redmagic\ galaxies than spectroscopic galaxies in our footprint.
Second, because the limited redshift range of our \redmagic\ \pzs\ means we can only measure part of the source redshift distribution directly with \wz, we use the \wz\ instead to calibrate shifts to the redshift distributions measured by \pz.
This calibration procedure is combined with \texttt{COSMOS} calibrations described in \citet{photoz} to obtain yet tighter constraints on the mean redshift of each source bin.

This paper is organised as follows.
In Section~\ref{sec:dataset} we present the various galaxy samples, flux measurement methods, and \pz\ algorithms used.
In Section~\ref{sec:methods} we briefly present the theory behind measuring redshift distributions with \wz\ and the way we calibrate \pz\ distributions with \wz.
In Section~\ref{sec:results} we present our calibrations of the redshift distributions.
In Section~\ref{sec:systematics} we compare the systematic errors we found in simulations in \gatti\ to their equivalent measurements in data.
In Section~\ref{sec:discussion} we compare our calibration with a calibration of the redshift distributions from \spec\ 30-band photometry described in \citet{photoz}.
Finally, in Section~\ref{sec:conclusions} we discuss future prospects for this method and present our conclusions.

\section{Data}
\label{sec:dataset}

The Dark Energy Survey is a 5000 square degree photometric survey that will image about 300 million galaxies in $grizY$ filters up to a redshift of $z = 1.4$ with the Dark Energy Camera~\citep[DECam,][]{Flaugher:2015aa}, a 570-megapixel camera built by the collaboration and stationed at the Cerro Tololo Inter-American Observatory (CTIO) 4-meter Blanco telescope. Here we use the DES Year 1 (Y1) data, which are based on observations taken between 31 August 2013 and 9 February 2014 during the first full season of the survey. We use the Y1 observations of the region overlapping with the South Pole Telescope \citep[SPT]{Carlstrom:2011aa} footprint.
These data are processed and extensively tested, resulting in a `Gold' catalog of objects \citep{y1gold}.

\subsection{Reference \redmagic\ Galaxies}

From the DES Y1 Gold Catalog we select a subset of objects using the \redmagic\ algorithm \citep{redmagicSV}. The algorithm aims to define a sample of luminous red galaxies above some minimal luminosity threshold. These galaxies are selected by fitting to a red sequence template that was calibrated using galaxy clusters selected by the redMaPPer  algorithm \citep{Rykoff:2014aa,Rykoff:2016aa}. These galaxies are selected to have a constant comoving number density of galaxies as a function of redshift, and have excellent photo-$z$'s, with an approximately Gaussian scatter of $\sigma_z / (1 + z) < 0.02$ \citep{redmagicSV}. The high quality of the \redmagic\ photometric redshifts makes them an appropriate sample for measuring \wz.

There are several variants of \redmagic\ we could use, based on the luminosity threshold and the input photometry (see Section~\ref{sec:dataset:fluxes}). We choose to use the `higher luminosity' sample, which has a luminosity threshold of $L > 1.5 L \star$, and which uses the `multi-object fitting' (\mof) photometry described in \citet{y1gold} and in Section~\ref{sec:dataset:fluxes} below. Further, when calculating correlations, we correct for systematic correlations between the object density and survey properties as detailed in \citet{wthetapaper}.

In \citetalias{xcorrtechnique}, the red sequence template measured in simulations by redMaPPer is redder than in DES Y1 data. Consequently, the maximum redshift in simulation over much of the footprint is $z_{\mathrm{max}} = 0.85$. For consistency with the analyses in the simulations from which we derive our errors, we also will use $z_{\mathrm{max}} = 0.85$ when we perform our analyses. Because our constraints here are derived from windowed means, the exclusion of $0.85 < z < 0.90$ redshift range produces negligible changes to our calibration.

The calibration of the \pz's of \redmagic\ galaxies via clustering with spectra is the subject of \citet{redmagicpz}.

\subsection{Source Galaxies}

After the `Gold' catalog is created, shape measurement algorithms are run on the galaxies to produce the shape catalogs that go into our cosmology analyses. The validation of these catalogs is the subject of \citet{shearcat}. It suffices for us to say that the primary catalog is the \texttt{METACALIBRATION} catalog \citep{metacal1, metacal2}, and we also calibrate a second shape catalog \texttt{IM3SHAPE} \citep{im3shape}. These galaxies are placed into four redshift bins with edges $(0.2, 0.43, 0.63, 0.9, 1.3)$ based on the point estimate of a \pz\ algorithm (see Section~\ref{sec:dataset:photoz}).
We cannot constrain the source redshift distribution beyond $z_{\mathrm{max}}$ using the correlation method. We are reliant upon the \pz\ determination of $n(z)$ outside the \redmagic\ redshift range, particularly in the highest-redshift bin. It is shown in \citet{keypaper} that the uncertainties in this extrapolation do not significantly alter the Y1 cosmological inferences.

\subsection{Photometric Redshifts}
\label{sec:dataset:photoz}

For redshift bin assignment and redshift distribution evaluation we turn primarily to a modified version of the Bayesian Photometric Redshifts (\bpz) code \citep{Benitez:2000aa}. This method uses a set of interpolated model spectral templates to calculate the likelihood of a galaxy's photometry belonging to a given template at a given redshift via $\chi^2$ between the observed fluxes and those of the filters integrated over the model template. A prior is then applied to the likelihood which consists of the relative luminosity functions of the templates and the distribution of magnitude in the $i$-band as a function of redshift. The details for how the templates are generated, how the prior is calibrated, and how we have modified the code may be found in \citet{photoz}.

We also provide calibrations for the \pz's obtained using Directional Neighborhood Fitting (\dnf) of \citet{jvicente}. \dnf\ is a machine-learning algorithm which takes as a training sample a collection of galaxies whose spectroscopic redshifts are known. Based on this training sample, \dnf\ makes a predictive hyperplane to best fit the neighborhood about each target galaxy in flux space, which is then used to predict the redshift. Details about the training and validation of the Y1 \dnf\ predictions may also be found in \citet{photoz}.

\subsection{\Pz\ Distributions}
\label{sec:pz_redshifts}

In this section we briefly explain how we obtain tomographic redshift distributions from our \pz\ catalogs.
For every galaxy, the \pz\ method returns a probability distribution of the redshift given its measured fluxes, plus a point estimate that is used for bin assignment. For both \bpz\ and \dnf\ this point estimate is simply the mean of the distribution.
To obtain the redshift distribution of a specific tomographic bin, we then `stack' the \pz\ probabilities of the galaxies using weights $w_j$ assigned by the shape measurement algorithm:
\begin{equation}
    \nz(z) = \frac{\sum_{j \in \mathrm{Bin} \ i} w_{j} P_{j}(z) }{\sum_{j \in \mathrm{Bin} \ i} w_{j}} \ ,
    \label{eq:stacking}
\end{equation}
where $P_{j}(z)$ is the probability of a given galaxy having redshift $z$. These weights are also used when we count pairs. We note that other uses of shape catalogs can have different lensing weights. These naturally lead to different effective redshift distributions and necessitate separate calibrations.

In practice, for every galaxy we draw a single redshift sample from $P_{j}$ and then create a weighted histogram from that. Because we are using millions of galaxies in our tomographic bins, this is for all practical purposes equivalent to Equation~\eqref{eq:stacking}.

\subsection{`\texttt{MOF}' and `\texttt{METACAL}' Fluxes}
\label{sec:dataset:fluxes}

In the \fiducial\ sample we have two different measurements of a galaxy's flux. The `multi-object fitting' (\mof) fluxes are described in \citet{y1gold}. In brief, the \texttt{NGMIX}\footnote{https://github.com/esheldon/ngmix} code fits cutouts of DES Y1 galaxies to a highly constrained exponential$+$deVaucouleurs model convolved with each exposure's point-spread function. The fit is multi-epoch and multi-band, with common shape parameters across bands and a single free flux per band. The fit also accounts for flux from neighbor galaxies in an iterative fashion, subtracting the current estimate of neighboring galaxies' flux. In contrast, the \texttt{METACALIBRATION} catalog fits a simple elliptical Gaussian convolved with each exposure's point-spread function. The fit is again multi-epoch and multi-band, with a single free flux per band, but it is not multi-object: no neighbor flux subtraction occurs. We shall refer to these two flux types as `\mof' and `\metacal' respectively.

The tomographic bin assignment and the redshift distribution evaluation may be done with different \pz\ algorithms, or even the same algorithm but with different flux types. For example, the \fiducial\ sample uses the \bpz\ algorithm for both tasks, but assigns galaxies to tomographic bins based on \metacal\ fluxes, and evaluates redshift distributions with \mof\ fluxes. The redshift distributions derived from clustering depend on the shape catalog and bin assignment, but not on the redshift distribution estimated by the \pz\ algorithm.
Of most interest are the following combinations of shape catalog, bin assignment, and redshift evaluation: the fiducial sample of the \texttt{METACALIBRATION} catalog placed into redshift bins by \bpz\ run on \metacal\ fluxes and whose redshift distribution is evaluated by \bpz\ run on \mof\ fluxes; and \texttt{IM3SHAPE} binned and evaluated by \bpz\ run on \mof\ fluxes. In \citet{keypaper} these different samples are used to test the robustness of our analysis to different shape and redshift algorithms.
For each of these two shape catalogs we may easily change the algorithm and fluxes used to evaluate the redshift distribution, so we also present their calibrations.

Table~\ref{tab:desy1_full} lists the mean redshifts in each bin for each redshift distribution. They are grouped by shape catalog and \pz\ algorithm used for tomographic assignment. Each group evaluates the redshift distribution of the same set of galaxies. The task of our calibration procedure is to correct these redshift distributions to have the true mean redshift.

\begin{table*}
\centering
\begin{tabular}{| l | l | l || c | c | c |}
  \hline
  Catalog & Photo-$z$ Binning & Photo-$z$ Redshift & $\langle z^1 \rangle + \Delta z^1 \pm \sigma_{\Delta z^1}$ & $\langle z^2 \rangle + \Delta z^2 \pm \sigma_{\Delta z^2}$ & $\langle z^3 \rangle + \Delta z^3 \pm \sigma_{\Delta z^3}$ \\
&  &  & $0.2 < z < 0.43$ & $0.43 < z < 0.63$ & $0.63 < z < 0.9$ \\
\hline
  \hline
\rowcolor{Gray}
\texttt{METACAL} & \texttt{METACAL BPZ} & \texttt{MOF BPZ}      &  $0.378 + 0.007 \pm 0.026$ &  $0.515 - 0.023 \pm 0.017$ &  $0.740 + 0.003 \pm 0.014$ \\
\texttt{METACAL } & \texttt{METACAL BPZ} & \texttt{MOF DNF}     &  $0.422 - 0.055 \pm 0.014$ &  $0.535 - 0.047 \pm 0.014$ &  $0.747 - 0.001 \pm 0.019$ \\
\texttt{METACAL} & \texttt{METACAL BPZ} & \texttt{METACAL BPZ}  &  $0.359 + 0.008 \pm 0.026$ &  $0.527 - 0.050 \pm 0.017$ &  $0.750 - 0.016 \pm 0.014$ \\
\texttt{METACAL} & \texttt{METACAL BPZ} & \texttt{METACAL DNF}  &  $0.421 - 0.051 \pm 0.014$ &  $0.538 - 0.053 \pm 0.014$ &  $0.747 - 0.002 \pm 0.019$ \\
 & & & & & \\
\rowcolor{Gray}
\texttt{IM3SHAPE} & \texttt{MOF BPZ} & \texttt{MOF BPZ}         &  $0.360 + 0.008 \pm 0.026$ &  $0.516 - 0.031 \pm 0.017$ &  $0.750 - 0.010 \pm 0.014$ \\
\texttt{IM3SHAPE} & \texttt{MOF BPZ} & \texttt{MOF DNF}         &  $0.430 - 0.078 \pm 0.014$ &  $0.553 - 0.059 \pm 0.014$ &  $0.754 - 0.025 \pm 0.019$ \\
 & & & & & \\
\texttt{METACAL} & \texttt{METACAL DNF} & \texttt{MOF BPZ}      &  $0.387 - 0.007 \pm 0.026$ &  $0.490 + 0.008 \pm 0.017$ &  $0.747 + 0.017 \pm 0.014$ \\
\texttt{METACAL} & \texttt{METACAL DNF} & \texttt{MOF DNF}      &  $0.377 - 0.004 \pm 0.014$ &  $0.524 - 0.034 \pm 0.014$ &  $0.758 + 0.007 \pm 0.019$ \\
\texttt{METACAL} & \texttt{METACAL DNF} & \texttt{METACAL BPZ}  &  $0.393 - 0.033 \pm 0.026$ &  $0.505 - 0.005 \pm 0.017$ &  $0.767 + 0.011 \pm 0.014$ \\
\texttt{METACAL} & \texttt{METACAL DNF} & \texttt{METACAL DNF}  &  $0.371 + 0.003 \pm 0.014$ &  $0.528 - 0.037 \pm 0.014$ &  $0.760 + 0.005 \pm 0.019$ \\
  \hline
\end{tabular}
\caption{
Table of mean redshifts, calibrations, and errors on the calibrations for the first three tomographic bins ordered by shape catalog, \pz\ used in tomographic binning, and \pz\ used in estimating the redshift distribution. Each grouping by shape and binning measures the same underlying redshift distribution. The numbers are prior to calibration. Grey rows are samples used in \citet{keypaper} to check the robustness of cosmological measurements.
}
\label{tab:desy1_full}
\end{table*}

\subsection{Simulations and Estimation of Systematic Error}

Our systematic errors are estimated from the \texttt{Buzzard-v1.1} simulation and are described in more detail in \citetalias{xcorrtechnique}. The simulation and creation of mock survey data are described in \citet{DeRose2017, Wechsler2017, simspaper}, but we now provide a brief summary. Three $N$-body simulations are run using a modified version of \texttt{GADGET2} \citep{springel2005} called \texttt{L-GADGET2}. The boxes range in size from one to four Gpc/$h$ on a side. \texttt{ROCKSTAR} \citep{Behroozi2013} identifies halos of dark matter, and galaxies are added to the simulations via the \texttt{ADDGALS} algorithm \citep{Wechsler2017}. Spectral energy distributions (SEDs) from the Sloan Digital Sky Survey (SDSS) DR7 \citep{Cooper2011} are assigned to galaxies based on local environmental density, which are then integrated through the Dark Energy Camera filters to generate \textit{grizY} magnitudes. Galaxy positions, shapes, and magnitudes are lensed using the \texttt{CALCLENS} ray-tracing code \citep{Becker2013}, and are cut to lie in the DES Y1 footprint using the DES Y1 depth maps \citep{Rykoff15,y1gold}.

From the simulated galaxies we construct our source and reference galaxy distributions. The redMaPPer and \redmagic\ algorithms are run to produce reference galaxies. The bias and scatter mimic observed \redmagic\ performance from spectroscopic SDSS samples. As described above, the red sequence in this simulation is redder than in DES Y1 data, leading to a maximum redshift of $z_{\mathrm{max}} = 0.85$ over much of the simulation footprint. This arises from the way SEDs are assigned to galaxies, and will be fixed in future iterations of these simulations. We find that all other relevant properties of the simulated \redmagic\ catalogs -- for example, clustering and number density -- are sufficiently accurate for our calibrations.
In order to create our simulation source galaxies, we cut on size and flux to mimic the \texttt{METACALIBRATION} sample. Both \pz\ algorithms are run on these galaxies to produce the redshift distributions. The shapes of the redshift distributions as estimated by the \pz\ algorithms matches well-enough what we observe in data, although imperfections in galaxy SED modeling lead to small differences.

By having a simulated dataset that mimics our data we are able to estimate systematic errors arising from our calibration by calibrating the simulated catalogs and comparing to the true redshifts. The simulated catalogs are quite good, but they are not perfect. For example, the distribution of SEDs with redshift, luminosity, and environment may be different in the simulation than in our universe, but the magnitude of the scatter should be reliable. In Section~\ref{sec:systematics} we will compare our results against the simulation results found in \citetalias{xcorrtechnique}.

\section{Methods}
\label{sec:methods}

\subsection{Determining the \Wz}

We outline the theory in this section and present our estimator. The
approach presented here is based on \citet{Menard:2013aa}, \citet{Schmidt:2013ac}, and \citet{Davis:2017aa}, with some modifications. We refer interested readers to \gatti\ for further details.

Our goal is to measure the redshift distributions of the \fiducial\ sample by measuring the cross-correlation of its angular positions with a reference sample which has well-measured redshifts. We shall refer to our reference sample by the superscript $r$ and our source galaxy sample by the superscript $u$, for `reference' and `unknown' respectively. The angular cross-correlation between two samples $n^u(z)$ and $n^r(z)$ with (linear) biases $b^u$ and $b^r$ is
\begin{align}
    w^{ur} = \int dR dz' dz''\ W(R) & n^u (z') n^r (z'') \ \times \nonumber \\
                                    & b^u(R, z') b^r(R, z'') \xi(R, z', z'') \,
\label{eq:w_definition}
\end{align}
where $\xi$ is the matter--matter correlation function $\avg{\delta \delta}(R, z', z'')$, and we allow the galaxy clustering bias $b$ to have both redshift and scale dependence for some separation $R$ and redshift $z$. We choose to integrate over \textit{comoving} scales $R = (1 + z) D_{A}(z) \theta$, from $\Rmin$ to $\Rmax$, which we choose to be 500 to 1500 kpc \citepalias{xcorrtechnique}. We explore the effects of varying these scales in Section~\ref{sec:scales}. Following \citet{Schmidt:2013ac}, we weight this integration with $W(R) \propto R^{-1}$ and the normalisation defined such that the weight function integrates to one over our chosen scales.
If we consider the case where we slice the reference sample into thin redshift bins such that $n^r(z'') = \delta(z - z'')$ for a slice centred at redshift\footnote{In practice, our reference sample has redshift errors of order $\sigma_z/(1 + z) < 0.02$. We slice our reference sample into bins of width $0.02$, although we note that our results are largely independent of reference bin size \citepalias{xcorrtechnique}.} $z$ and if we use the fact that $\xi$ drops off quickly with comoving separation, such that $\xi(R, z', z'') = \xi(R, z', z'') \delta(z' - z'')$, we may pull out the redshift distribution $n^u$, upgrade $w^{ur}$ to a function with a dependence on redshift $z$, and arrive at
\begin{equation}
w^{ur}(z) = n^u (z) \int_{\Rmin}^{\Rmax} dR \ W(R) b^u(R, z) b^r(R, z) \xi(R, z, z) \ .
\end{equation}
We define the function $f(z)$ to be the weighted integral of the product of the bias terms and the projected matter-matter correlation function:
\begin{equation}
    f(z) = \int_{\Rmin}^{\Rmax} dR \ W(R) b^u(R, z) b^r(R, z) \xi(R, z, z) \ .
\end{equation}
The unknown function $f(z)$ characterises the (possibly non-linear) growth in the correlation function and/or possibly evolving non-linearities in the correlation function. For the purposes of this paper, we have little power in constraining the value of $f(z)$, and instead will use simulations to characterise its impact in our calibration.
We are finally left with
\begin{equation}
w^{ur}(z) = f(z) n^u (z) \ ,
\label{eq:w_final}
\end{equation}
which relates the weighted average of the angular correlation function to the redshift distribution of the sample of interest.

We measure the angular correlation function by counting the number of pairs between our unknown and reference data $\Du\Dr$ separated over a range of comoving separations from $\Rmin$ to $\Rmax$.
Given the true angular correlation $w^{ur}$, the number of unknown--reference
pairs is:
\begin{align}
    \Du\Dr(z) = \bar{n}^u \bar{n}^r \int_{\Rmin}^{\Rmax} dR \int dx dx' \ & W(R) \left[1+w^{ur}(R, z)\right] \ \times \nonumber \\
    & S(x) S(x') \Theta(x,x',R) \ ,
    \label{eq:paircounts}
\end{align}
where $\bar{n}^u$ and $\bar{n}^r$ are the number densities of the unknown and reference samples,
$S(x)$ is the survey window function, such that $S(x) = 1$ if sky coordinate $x$ is in the survey and 0 otherwise, $\Theta(x,x',R)$ is a step function that is 1 if the distance between points $x$ and $x'$ is $R$ and 0 otherwise, and $dx$ and $dx'$ are double integrals over the sky.
In practice, the terms representing the area are estimated using `randoms,' or catalogs of random points placed on the sky that mimic where a galaxy \textit{could} have been observed. We may then measure their pairs and recover Equation~\eqref{eq:paircounts}, only with $w^{ur}=0$. However, it is difficult to estimate the selection function of galaxies in shape catalogs. The reasons a galaxy may or may not pass shape measurement quality flags are numerous and can depend on location in the sky in complicated and opaque manners. Consequently, there are no randoms for the shape catalogs, and so we cannot choose the common Landy-Szalay estimator \citep{Landy:1993aa}. Instead, the estimator we use is analogous to $\hat{w}_2$ in \citet{Landy:1993aa}:
\begin{equation}
    \hat{w}^{ur}(z) = \frac{\DD}{\DRu} \frac{N_{\Rr}}{N_{\Dr}} \left( z \right) - 1 \ ,
\label{eq:natural_noRu}
\end{equation}
where $N_{\Rr}$ is the total number of reference randoms and $N_{\Dr}$ is the total number of reference galaxies.
In the limit of infinitely large random catalogs its expectation value is
\begin{equation}
\langle \hat{w}^{ur} \rangle (z) = f(z) n^u (z) \ .
\end{equation}
Thus we have shown that we can use angular cross-correlations to obtain measurements of the redshift distribution, and that our ability to calibrate a redshift distribution depends upon our understanding of the evolution in redshift of the matter--matter correlation function and the biases of both the reference and unknown samples.

\subsection{Calibrating Photometric Redshift Distributions with \Wz}
\label{sec:calibration}

We use the cross-correlation measurements of the redshift distribution to calibrate the bias in photometric redshift estimates as described in \gatti.  In brief, for the $i$-th redshift bin with a photometrically estimated redshift distribution $\nz(z)$, we solve for the bias $\Delta z^i$ such that $\nz(z-\Delta z^i)$ has the same mean source redshift over the same window as the \wz\ $\hat{w}^{ir}$:\footnote{We adopt the convention that a positive $\Delta z^i$ means that galaxies are \textit{farther} away than our initial measurements indicated.}
\begin{equation}
    \frac{\int_{\zmin^i}^{\zmax^i} dz \ z \nz(z -  \Delta z^i)}{\int_{\zmin^i}^{\zmax^i} dz \ \nz(z - \Delta z^i)} = \frac{\int_{\zmin^i}^{\zmax^i} dz \ z \hat{w}^{ir}(z)}{\int_{\zmin^i}^{\zmax^i} dz \ \hat{w}^{ir}(z)} \ .
\label{eq:averaging}
\end{equation}
In Equation~\eqref{eq:averaging} we implicitly assume that $f(z)$ is constant. We check in \gatti\ with simulations how much this assumption biases our calibration, and find that while it is one of our dominant systematics, it is sufficiently small for our DES Y1 analysis.
We must also decide upon a minimum and maximum redshift for evaluating the mean redshift. We want sufficient range to get a good measure of the mean, but if the range is too broad, then we include noisy points in the tails. We decide upon the following algorithm for determining the range of redshifts used in the fit based on comparisons with simulations in \gatti:
\begin{itemize}
    \item{Calculate the mean and standard deviation of the \wz\ signal:
\begin{align}
    \label{eq:mean}
    \bar{z}^i &= \frac{\int dz \ z \hat{w}^{ir} (z)}{\int dz \ \hat{w}^{ir} (z)} \\
    (\sigma^i)^2 &= \frac{\int dz \ (z - \bar{z}^i)^2 \hat{w}^{ir} (z)}{\int dz \ \hat{w}^{ir} (z)} \ .
    \label{eq:width}
\end{align}
We integrate the \wz\ signal over $0.15 < z < 0.85$ in the first two tomographic bins, and $0.40 < z < 0.85$ in the third bin. In the third bin we cut $z < 0.40$ for the reason that we found systematically negative \wz\ signal ($\hat{w}^{ir} < 0$) at lower redshifts; reference galaxies at low redshifts are \textit{anti}-correlated with the source galaxy sample. \citetalias{xcorrtechnique} showed that low-redshift reference galaxies cross-correlating with high-redshift source galaxies could be anti-correlated due to magnification. Modeling magnification is outside the scope of this paper. Furthermore, all redshift codes indicate that there are very few galaxies in the third tomographic bin with $z < 0.40$.
Because the negative $\hat{w}^{ir}$ biases our mean and widths in Equations~\eqref{eq:mean} and \eqref{eq:width}, we exclude these points.
}
    \item{Choose a two sigma cut about the mean: $\zmin^i = \bar{z}^i - 2 \sigma^i$ and $\zmax^i = \bar{z}^i + 2 \sigma^i$.}
\end{itemize}
We will return to the impact of the choice of $2 \sigma^i$ in Section~\ref{sec:sigma}.

There are many ancillary choices to be made in this analysis. For example, we could have chosen a different range of scales, a different redshift range, or a different luminosity threshold for our reference \redmagic\ galaxies. Where possible, we tried, in an attempt to ensure our methodology is somewhat `blinded,' to decide our fiducial methodology purely by considering the impact of our choices on simulations. Thus, we find in \gatti\ with simulations that our calibration procedure performed best if we used the method from \citet{Schmidt:2013ac} for determining the redshift distribution, and we decide through these simulations that we would not apply an auto-correlation bias correction. Where possible, we check these choices on data. These checks are outlined in Section~\ref{sec:systematics}.

We also use simulations to estimate systematic errors in our calibration procedure. In \gatti, we find that our dominant systematic errors are (1) the bias evolution in our source galaxy sample caused by the tomographic binning procedure; and (2) mismatch between the shapes of the \pz\ and \wz\ distribution, including errors arising from asymmetry in the outlier galaxies and from our decision to use windowed means to calibrate shifts to the photometric redshift distribution.

The total systematic uncertainty is estimated as the sum in quadrature of all uncertainties, and varies between 0.014 and 0.026 depending on the tomographic bin in question.
The statistical precision of this measurement of the $\Delta z^i$'s is $\sim \! 0.005$. Thus we are dominated by systematic errors, but at an acceptable level for our cosmological constraints.
When we quote constraints on $\Delta z^i$, our errors are the quadrature sum of our statistical precision and these systematic errors.

\section{Results}
\label{sec:results}

\begin{figure*}
    \begin{center}
        \includegraphics[width=\textwidth]{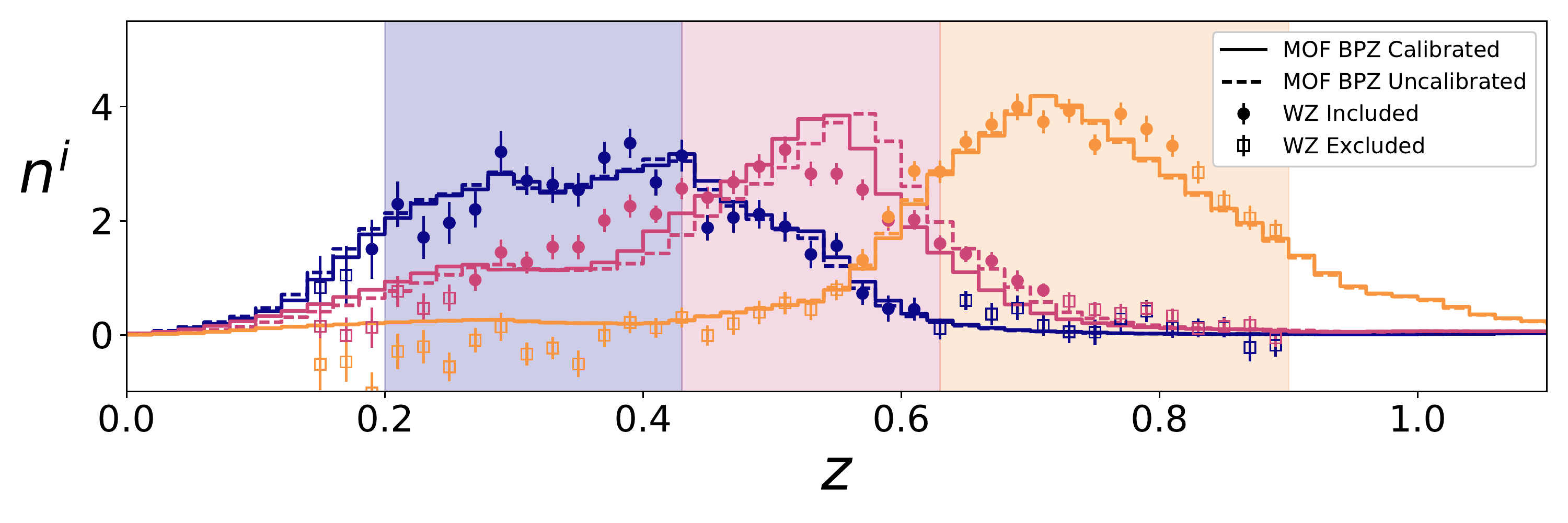}
    \end{center}
    \caption{
        The redshift distributions of the fiducial \texttt{METACALIBRATION} sample as measured by the uncalibrated and calibrated \bpz\ algorithm (dashed and solid lines) and by \wz\ (points). Galaxies are placed into each tomographic bin according to their mean \bpz\ estimate (background colors). Each color corresponds to a different tomographic bin.
    The \pz\ distributions are calibrated by the \wz\ points (circles) as described in Section~\ref{sec:calibration}, while the unfilled squares are excluded by the $\pm 2 \sigma$ window. The values of the mean redshift calibrations shown may be found in Table~\ref{tab:desy1_full}.}
    \label{fig:comparison}
\end{figure*}

We present our \wz\ distributions and the calibrated \pz\ distribution for the \fiducial\ sample in Figure~\ref{fig:comparison}. As we described in Section~\ref{sec:dataset}, galaxies are assigned to redshift bins by point estimates (mean of the individual galaxy probability distributions) from \bpz\ on \metacal\ fluxes. Measurements of the tomographic bins are denoted by their color. The points represent our \wz\ measurements, and the empty squares are points excluded by our $\bar{z}^i \pm 2 \sigma^i$ cut. The dashed lines are the original redshift distributions from stacking the \pz\ signal as described in Section~\ref{sec:pz_redshifts}. The solid lines are the calibrated \pz\ signal, $\nz(z - \Delta z^i)$, where $\Delta z^i$ is found by comparing with \wz\ as described in Section~\ref{sec:calibration}. In the first and third tomographic bins, the \wz\ distribution and the \pz\ distribution agree remarkably well, however the second tomographic bin features significant disagreement between the two redshift distributions. We reiterate, however, that for constraining cosmology, the most important quantity relating to the redshift distribution is its mean \citep{shearcorr}.

The constraints from our calibrations are presented in Table~\ref{tab:desy1_full}. As is evident from Figure~\ref{fig:comparison} and Table~\ref{tab:desy1_full}, the shifts are quite small by eye. Indeed, for the \fiducial\ sample the \wz\ measurements alone indicate that the shifts derived for \bpz\ are consistent with zero. The biggest corrections happen in the second tomographic bin, where the shapes of the \pz\ and \wz\ distributions are most discrepant.

\subsection{Other \Pz\ Algorithms and Shape Samples}
\label{sec:photozs}

In the creation of the \pz\ distributions, a \pz\ algorithm is used twice: first, in the assignment of galaxies to tomographic bins; second, in the measurement of the redshift distribution of a particular tomographic bin. In this section we present results for varying the \pz\ algorithm of these two parts.

\begin{figure*}
    \begin{center}
        \includegraphics[width=\textwidth]{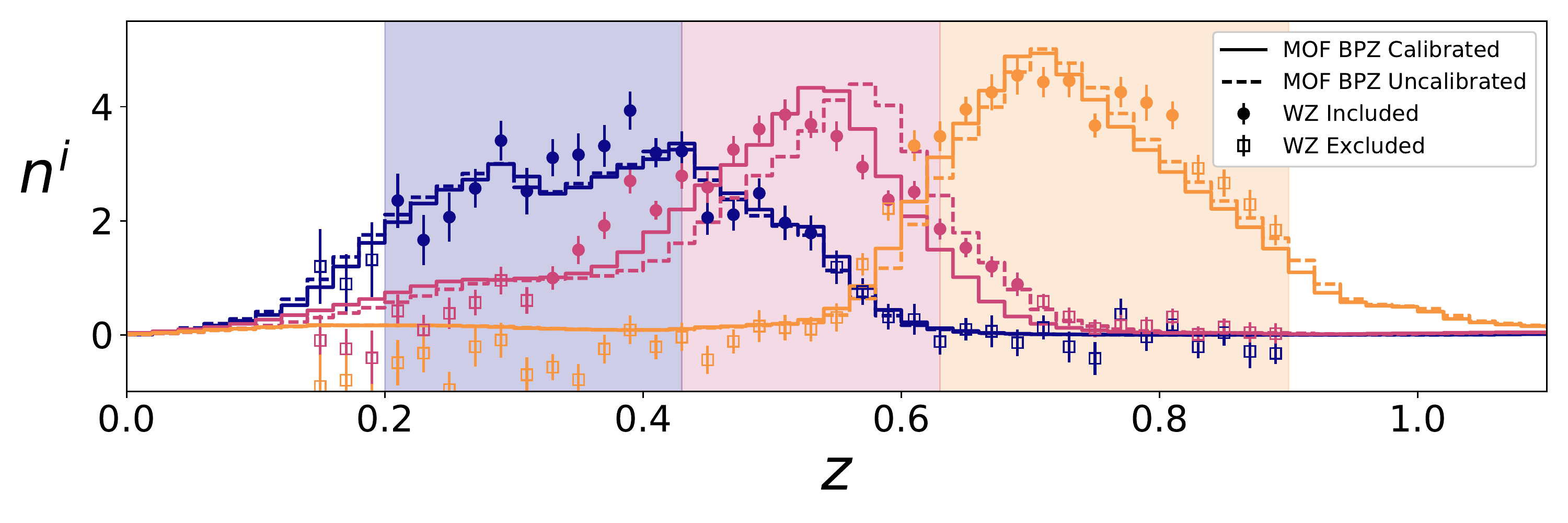}
    \end{center}
    \caption{
       The redshift distributions of the \imshape\ sample as measured by the uncalibrated and calibrated \bpz\ algorithm (dashed and solid lines) and by \wz\ (points). Galaxies are placed into each tomographic bin according to their mean \bpz\ estimate (background colors). Each color corresponds to a different tomographic bin.
    The \pz\ distributions are calibrated by the \wz\ points (circles) as described in Section~\ref{sec:calibration}, while the unfilled squares are excluded by the $\pm 2 \sigma$ window. The values of the mean redshift calibrations shown may be found in Table~\ref{tab:desy1_full}.
}
    \label{fig:dndz_im3shape}
\end{figure*}

\begin{figure*}
    \begin{center}
        \includegraphics[width=\textwidth]{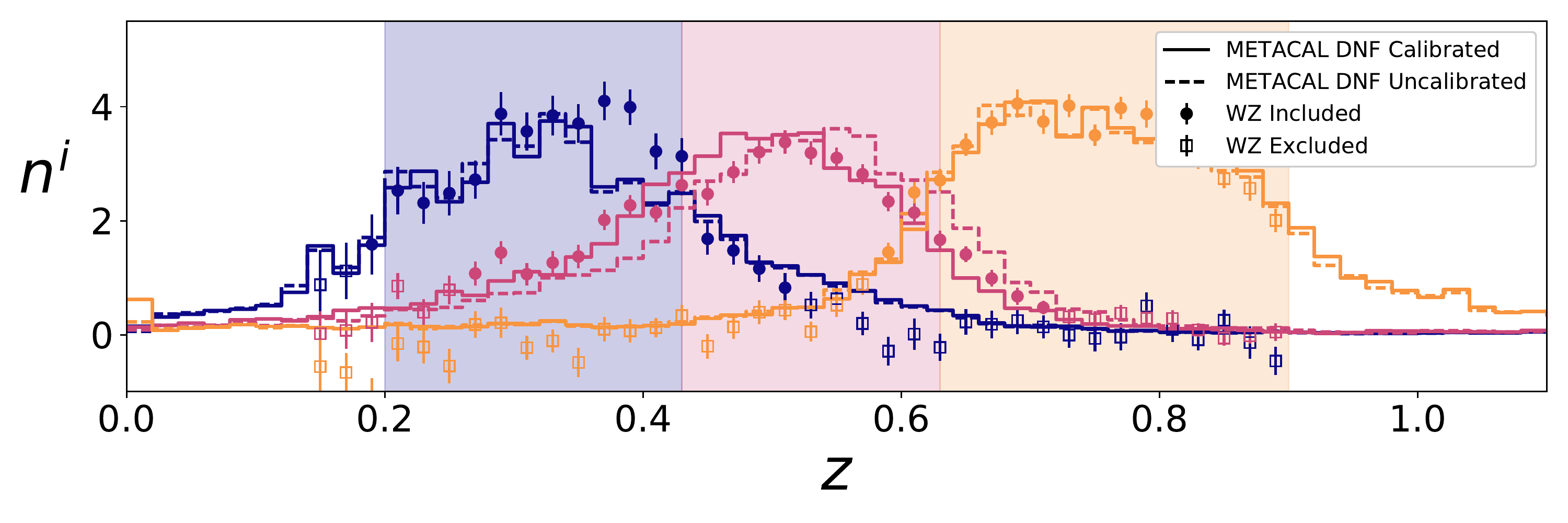}
    \end{center}
    \caption{The redshift distributions of the \metacal\ sample as measured by the uncalibrated and calibrated \dnf\ algorithm (dashed and solid lines) and by \wz\ (points). Galaxies are placed into each tomographic bin according to their mean \metacal\ \dnf\ estimate (background colors). Each color corresponds to a different tomographic bin.
    The \pz\ distributions are calibrated by the \wz\ points (circles) as described in Section~\ref{sec:calibration}, while the unfilled squares are excluded by the $\pm 2 \sigma$ window. The values of the mean redshift calibrations shown may be found in Table~\ref{tab:desy1_full}.
    Unlike Figure~\ref{fig:comparison}, galaxies are placed into tomographic bins based on the \dnf\ \pz\ algorithm.}
    \label{fig:dndz_dnf}
\end{figure*}

We examine a different shape catalog. We take the \imshape\ catalog, and bin and evaluate the redshift distributions based on the \bpz\ algorithm.
We show as solid lines the corrected redshift distribution after calibration from \wz.
These results are shown in Figure~\ref{fig:dndz_im3shape}. We also look at using \metacal\ \dnf\ to assign galaxies to redshift bins and evaluate redshift distributions, after which we repeat the same exercise. These results are presented in Figure~\ref{fig:dndz_dnf}.
The Figures show that the shifts are often quite small, and that the redshift distributions measured by \wz\ agree in the broad features with most of the \pz\ redshift distributions. We may also vary the \pz\ algorithm used to evaluate the redshift distribution of an ensemble of galaxies. We collect all of these results into Table~\ref{tab:desy1_full}.

After applying the calibrations the resultant calibrated means will still not be exactly the same because the matching is performed within a window about the mean of the \wz\ and not the full \pz\ distribution. Recall from Section~\ref{sec:calibration} that the windowing was chosen to mitigate the impact of noisy \wz\ tails. \citetalias{xcorrtechnique} estimates the impact of this effect on our calibrations to be at the $\sim0.01$ level or less, which is incorporated into our total systematic error. We can check the impact in data by looking at the dispersion in calibration of different \pz\ algorithms run on the same set of galaxies. We find that our calibrations on the same sets of galaxies (but different \pz\ distributions) have about this level of dispersion; it is not unreasonable that differences in shapes of the redshift distributions affect our calibrations at the $\sim0.01$ level.

We stress, however, that differences in shape, while interesting from many perspectives (\wz\ algorithm validation, \pz\ algorithm performance, galaxy evolution), do not significantly impact our cosmology analysis. As long as the mean is properly calibrated, which particular redshift distribution we use has a subdominant impact on cosmological constraints from our data \citep{shearcorr}.


\section{Systematic Errors}
\label{sec:systematics}

Many of the systematic error estimates made in \gatti\ can be verified on data. In this section, we check four of the sources: the choice of scales, the range of \wz\ signal used in the mean calibration, the bias evolution of the reference \redmagic\ sample, and finally our use of weights from systematics maps in our \redmagic\ galaxies and randoms.

\subsection{Dependence on Scale}
\label{sec:scales}

The choice of scale is an important decision in the \wz\ analysis. Larger scales tend to have poor signal-to-noise, while smaller scales are more likely to suffer from non-linear bias. In \gatti, we found with simulations that the \wz\ calibration was largely scale-independent, with a decrease in signal-to-noise at larger scales. We expected that on data the smaller scales would become suspect because of issues like deblending which were not modeled in the simulations. Avoiding these issues lead us in \gatti\ to decide on a fiducial range of scales of 500-1500 kpc. Variations in calibrations due to choice of scale was \textit{not} included as a systematic error.

\begin{figure*}
    \begin{center}
        \includegraphics[width=\textwidth]{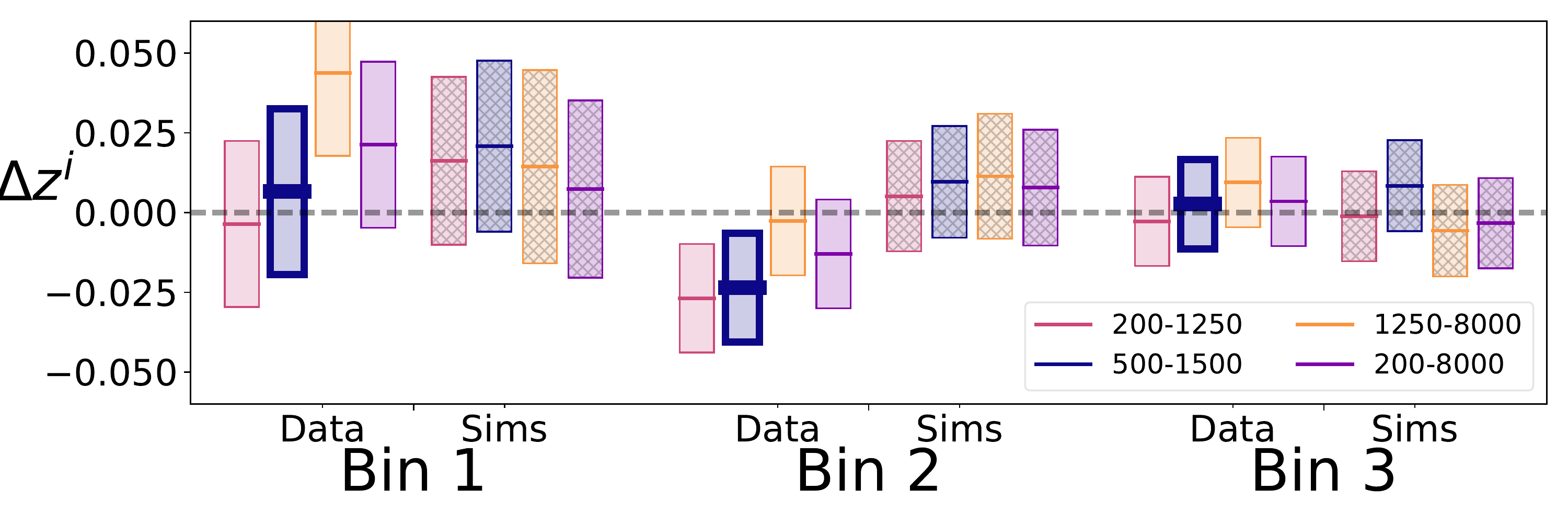}
    \end{center}
    \caption{
        Dependence of calibration on projected scales used in measuring the \wz\ signal in data (solid colors) and in simulations (hashed) for the three tomographic bins of the \fiducial\ sample. Our fiducial choice of scales is 500-1500 kpc (bold blue bars). The other scales used are 200-1250 kpc (pink), 1250-8000 kpc (yellow), and 200-8000 kpc (purple). Results are grouped by tomographic bin, with results on data on the left with solid bars and with results on simulations on the right with hashed bars. Errors are statistical and systematic added in quadrature. Because the redshift distributions differ between data and simulation, calibrated $\Delta z^i$ are different. The larger scales have appreciably different calibration $\Delta z^i$ than the other scale choices.
}
    \label{fig:scales}
\end{figure*}

\begin{figure*}
    \begin{center}
        \includegraphics[width=\textwidth]{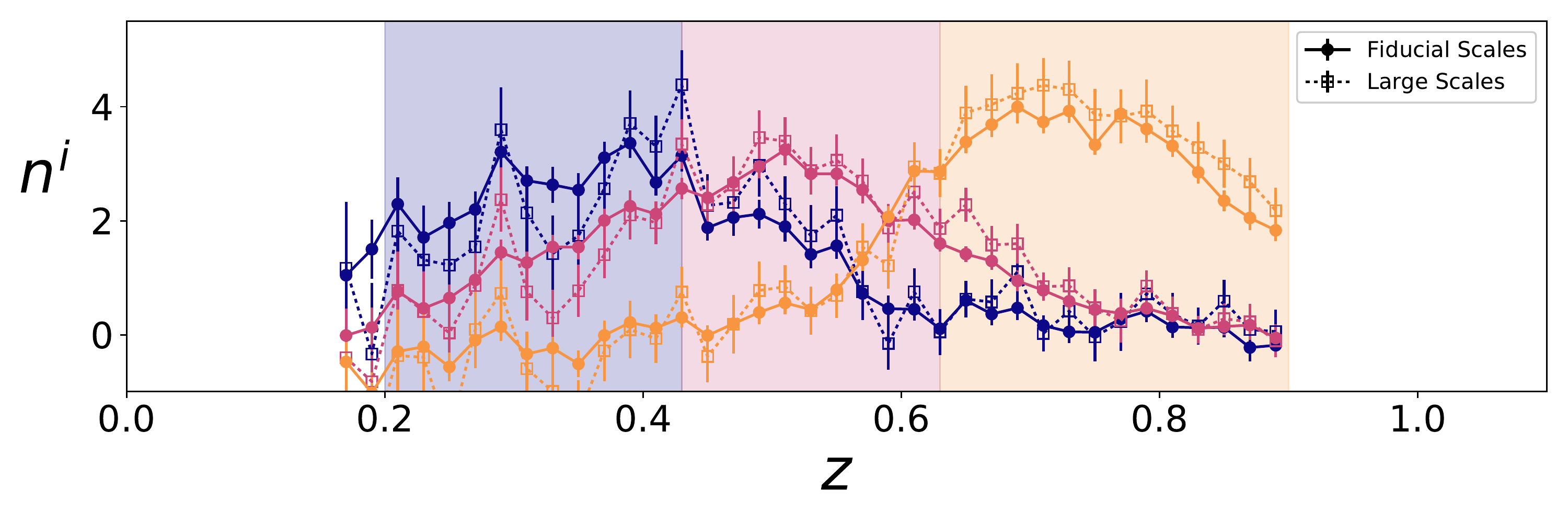}
    \end{center}
    \caption{
	Comparison of \wz\ signal from fiducial scales (500-1500 kpc; solid lines) and large scales (1250-8000 kpc; dotted lines) on the \fiducial\ sample. In addition to a tilt in the \wz\ from large scales relative to the fiducial scales, there is noticeable correlation between the redshift bins in the large-scales \wz\ which is not observed on the smaller scales. The differences observed here between the two scales correspond to the $\sim 0.02-0.03$ shifts in calibrations plotted in Figure~\ref{fig:scales}.
}
    \label{fig:scales_nz}
\end{figure*}

We are able to test the accuracy of this assumption in data by repeating our calibration over multiple scale ranges and comparing the trend with our work on simulations in \gatti.
We caution that the $\Delta z^i$ measured in simulations does not have to be the same as in the DES Y1 sample, as spectral energy distributions used to generate galaxies in the simulation do not precisely match the spectral energy distributions of the galaxies in our final data sample. Improved simulations will be used for future analyses, but for the purposes of this paper we only need differential measurements (\textit{e.g.} of how the mean redshift shifts with different ranges of scales used), and therefore the current simulations are adequate.
In Figure~\ref{fig:scales} we show the value of the redshift calibration, $\Delta z^i$, as a function of the range of scales used in the analysis. For each redshift bin we show a range of scales probed: 200-1250 kpc, 500-1500 kpc, 1250-8000 kpc, 200-8000 kpc. The left bars are the scale variation in data, while the hatched bars on the right are the results run on simulations. We note that 500-1500 kpc is our fiducial choice of scales.
We find that calibrations using only larger scales (1250-8000 kpc) shift the calibration by $\sim0.03$ and $\sim0.02$ in the first and second redshift bins. We observe that the large-scale \wz\ are tilted relative to the fiducial scales. As can also be seen in Figure~\ref{fig:scales_nz}, the large-scale \wz\ have stronger correlations between redshift bins, and are also significantly noisier.
This noisiness reinforces our decision to use 500-1500 kpc as our fiducial choice of scales.

\subsection{Sensitivity to Choice of Redshift Window}
\label{sec:sigma}

When choosing to calibrate the redshift distributions by matching the means of the \wz\ and \pz\ signals, we must be clear about the range of redshifts used. The \wz\ measurements do not span the full range of redshift space, and low-signal tails can exhibit unphysical behavior like negative redshift distributions ($\hat{w}^{ir} < 0$), and unmodelled behavior from lensing magnification more strongly affect measurements of the mean redshift. We decided to evaluate the means of the redshift distributions in the space within two standard deviations about the mean of the \wz\ signal. \gatti\ characterises the uncertainty introduced by choosing two standard deviations by comparing the calibrations obtained by using 1.5 and 2.5 standard deviations. With these values, we found that we typically introduced an uncertainty of order 0.005 but in certain cases found uncertainties as high as 0.015. These uncertainties are included in our systematic error budget.

\begin{figure}
    \begin{center}
        \includegraphics[width=0.5\textwidth]{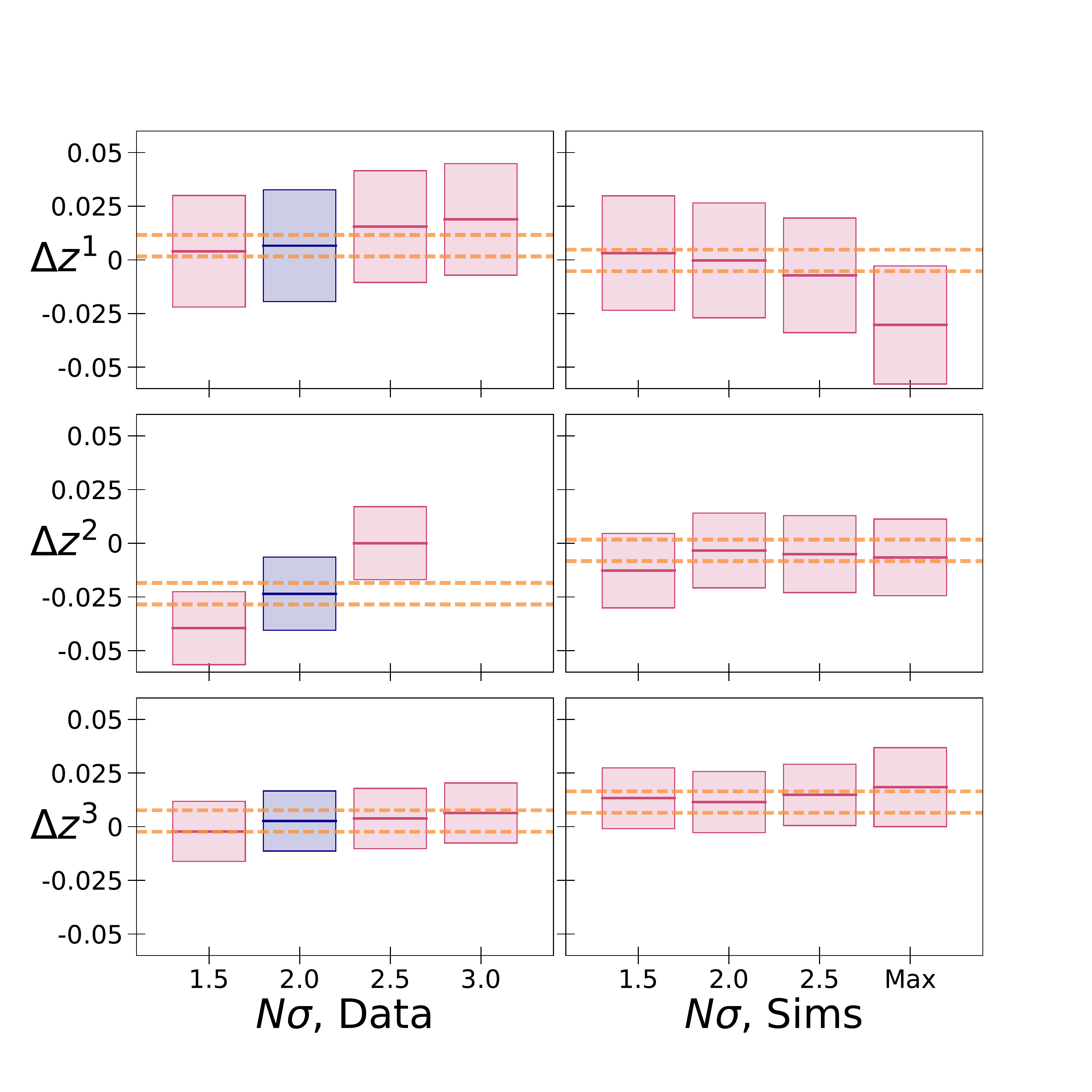}
    \end{center}
    \caption{Dependence of calibration on width of clustering redshift distribution considered in the calibration for the three tomographic bins in the \fiducial\ sample. The blue bar corresponds to the scale used in our analysis, and the orange dashed lines are that central value $\pm 0.005$, the approximate size of changing $N$ from 2 to 1.5 or 2.5 as measured in simulations \citepalias{xcorrtechnique}. The effects of changing $N$ in simulations are shown on the right plots, and are the same values as shown in Figure~5 from \citetalias{xcorrtechnique}. The `Max' column shows the measurements in simulations for $N = (3.5, 3.0, 5.0)$ for the three tomographic bins.
Errors are statistics and systematics added in quadrature. When considering only a small range of the \wz\ signal, skewness and shape discrepancies can bias the calibration. In contrast, considering the full redshift distributions can introduce biases from using noisy tails to perform the calibration. In the second tomographic bin for data, considering larger widths does not include any more points because we already use all points at $N = 2.5$. We choose to consider two standard deviations about the \wz\ signal as a reasonable compromise.
}
    \label{fig:sigmas}
\end{figure}

We check if this remains so in data. We illustrate the impact of different ranges of redshifts with Figure~\ref{fig:sigmas}. Our fiducial separation, $\bar{z}^i \pm 2 \sigma^i$, is plotted in blue, while other choices are shown in pink. We show the results from simulations on the right side. As we go to larger numbers of standard deviations, we reach our full \wz\ distribution, and so the calibration levels out for some tomographic bins.
The \wz\ distributions in real data tend to have broader and more skewed tails than in simulations. A consequence of this is that our calibrations do exhibit a stronger dependence on redshift range than in simulations, particularly in the second tomographic bin, where we find that going from 2 standard deviations to 2.5 changes our $\Delta z^i$ constraint by $\sim 0.025$. \gatti\ also find that this tomographic bin behaves worst, finding a shift of $0.012$. This is still a factor of two smaller than what we observe in data, but we caution that this second redshift bin is also the most discrepant between simulation and observation. The estimated redshift distribution in simulations is considerably narrower, and we do not find a low-redshift peak in the DES Y1 \wz\ distribution. Finally, 2.5 standard deviations includes all measured \wz, including the noisy tails that we sought to cut using the windowing procedure.

\subsection{Sensitivity to \redmagic\ Catalog and Weights}

We have some freedom in the particular reference catalog we use. While we used the \redmagic\ `higher luminosity' sample ($L > 1.5 L \star$), we also could have used the `high density' ($L > 0.5 L \star$) or `high luminosity' ($L > L \star$) samples,\footnote{These samples correspond to galaxy samples with lower luminosity thresholds but higher comoving densities, which also means their maximum redshifts are lower.} or even a combination of them. In the \fiducial\ cosmology analyses, a `combined' sample of \redmagic\ galaxies is used as lenses, where the `high density' sample is used over the redshift range $0.15 < z < 0.6$, the `high luminosity' over $0.6 < z < 0.75$, and the `higher luminosity' over $0.75 < z < 0.90$ \citep{wthetapaper, gglpaper}.
\gatti\ finds that the evolution of clustering bias is a dominant systematic, mostly from the \pz\ algorithm systematically identifying certain types of low redshift galaxies as high redshift, or vice versa, and then selecting galaxies into tomographic bins based on that estimated redshift.
For example, a galaxy whose colors might lead a \pz\ algorithm to assign it a moderate redshift (the algorithm measures it to be a redshifted blue galaxy), but which is actually at a low redshift (a red galaxy) has different clustering properties than a redshifted blue galaxy.. By stitching different samples together, we may introduce bias evolution as we compare `high density' clustering with `higher luminosity'.
In simulations we found that the bias of the `combined' and `higher luminosity' samples were compatible within our errors, and that the $\Delta z^i$ measured from one varied from the other at about the $\sim 0.005$ level.
We decided to choose the single \redmagic\ `higher luminosity' sample in order to minimise the potential for systematic errors from bias evolution in our reference sample. In repeating our analysis with the `combined' sample, we find that $\Delta z^i$ shifts by about no more than 0.0025.

In addition to the different \redmagic\ catalogs, we also weighted our reference galaxies by the systematics in our survey. \citet{wthetapaper} derives these weights in order to remove spurious correlations with systematics like seeing and exposure time. This consideration of correlations with systematics has, to our knowledge, not been considered in \wz\ studies. We can repeat our exercise without the weights in the reference galaxy sample. We find that while the redshift distributions are affected at the $\sim\!10\%$ level, the means are only negligibly shifted. We keep the correction, but conclude that its impact on our calibration is minor.

\section{Comparison with \Spec\ Calibration}
\label{sec:discussion}

In this Section we briefly recapitulate the \spec\ redshift distribution calibration. Interested readers should turn to \citet{photoz} for more details. We perform our \spec\ redshift distribution calibration by turning to the COSMOS2015 catalog from \citet{Laigle}. This catalog provides photometry in 30 different bands spanning the ultraviolet to the infrared as well as probability distribution functions for the redshift of each galaxy based on this photometry from the \textsc{LePhare} template-fitting code \citep{LePhare1,LePhare2}.

Galaxies in the \spec\ footprint are reweighted to match the color-size distribution of our source galaxy samples. We run our \pz\ algorithms on these galaxies, using their outputs to place galaxies into tomographic bins and to measure their redshift distributions. Because we also have the high fidelity redshifts, we can compute a shift in the mean redshift between the \pz\ and \spec\ redshifts. This quantity should yield the same $\Delta z^i$ as we measure here. Like our clustering calibration, the \spec\ calibration estimates uncertainties such as cosmic variance from the small size of the \texttt{COSMOS} footprint with simulations.

\begin{figure*}
    \begin{center}
        \includegraphics[width=\textwidth]{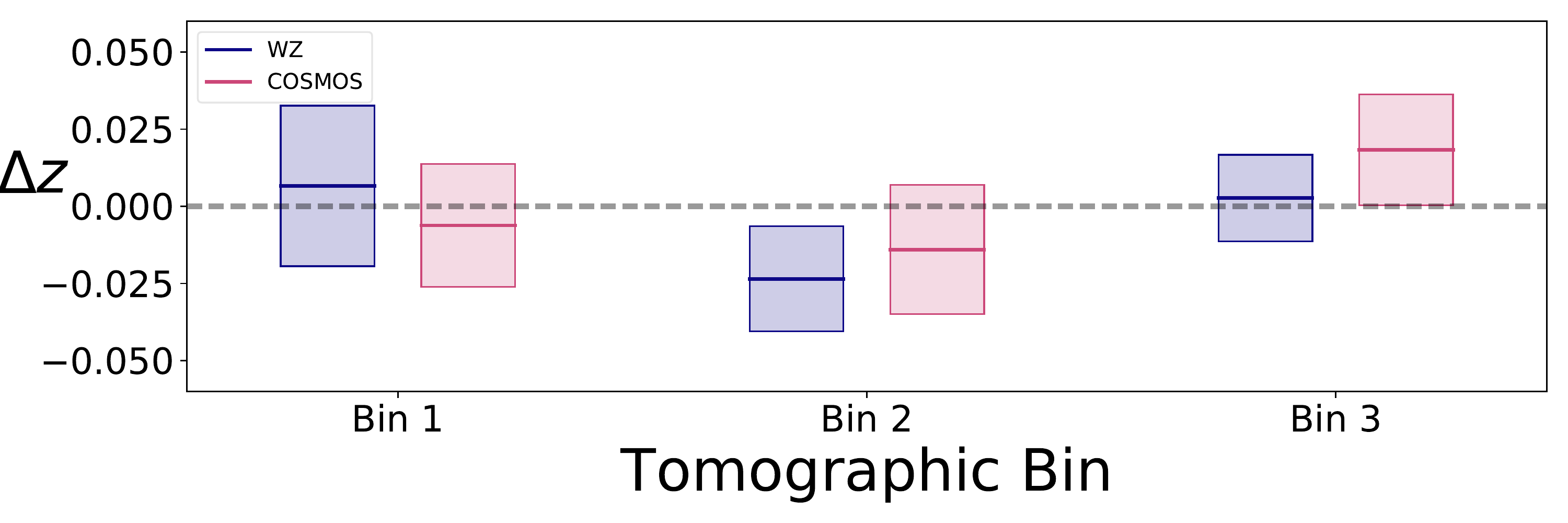}
    \end{center}
    \caption{
    Comparison of $\Delta z^i$ obtained from \wz, \spec, and the \combined\ calibrations for the \fiducial\ sample. The values may also be found in Table~\ref{tab:desy1}. The \wz\ measurements do not extend beyond $z=0.9$, and so a \wz\ calibration of the last tomographic bin is not performed. Despite using very different methods, the two modes of calibration agree remarkably well.
    }
    \label{fig:fiducial_cosmos}
\end{figure*}

A comparison of the calibrations for the fiducial sample can be found in Figure~\ref{fig:fiducial_cosmos}. The numerical values of several comparisons are listed in Table~\ref{tab:desy1}. In these plots, we show bars representing the one sigma constraints on $\Delta z^i$ as measured by both \wz\ and \spec\ in different tomographic bins. The calibrations are entirely compatible, despite being quite different procedures -- an indication of the robustness of our calibration procedure.

The \spec\ and \wz\ calibrations are combined in \citet{photoz} and lead to constraints on the mean redshifts of $\sim \pm 0.015$.

\begin{table*}
\centering
\begin{tabular}{| l | l | l | l || r | r |}
  \hline
Shape Catalog & Photo-$z$ & $z_{\mathrm{min}}$ & $z_{\mathrm{max}}$ & \WZ\ Correction & \SPEC\ Correction \\
\hline
\texttt{METACAL} & \texttt{METACAL/MOF BPZ} & 0.20 & 0.43 &  $+0.007 \pm 0.026$ &  $-0.006 \pm 0.020$ \\
 &  & 0.43 & 0.63 &  $-0.023 \pm 0.017$ &  $-0.014 \pm 0.021$  \\
 &  & 0.63 & 0.90 &  $+0.003 \pm 0.014$ &  $+0.018 \pm 0.018$  \\
 &  & 0.90 & 1.30 & - &  $-0.018 \pm 0.022$ \\
 & & & & & \\
\texttt{IM3SHAPE} & \texttt{MOF BPZ} & 0.20 & 0.43 &  $+0.008 \pm 0.026$ &  $+0.001 \pm 0.020$ \\
 &  & 0.43 & 0.63 &  $-0.031 \pm 0.017$ &  $-0.014 \pm 0.021$ \\
 &  & 0.63 & 0.90 &  $-0.010 \pm 0.014$ &  $+0.008 \pm 0.018$ \\
 &  & 0.90 & 1.30 & - &  $-0.057 \pm 0.022$ \\
 & & & & & \\
\texttt{METACAL} & \texttt{METACAL DNF} & 0.20 & 0.43 &  $+0.003 \pm 0.014$ &  $-0.024 \pm 0.017$ \\
 &  & 0.43 & 0.63 &  $-0.037 \pm 0.014$ &  $-0.042 \pm 0.021$ \\
 &  & 0.63 & 0.90 &  $+0.005 \pm 0.019$ &  $+0.006 \pm 0.021$ \\
 &  & 0.90 & 1.30 & - &  $+0.037 \pm 0.020$ \\
  \hline
\end{tabular}
\caption{
    Table of redshift distribution calibrations from \wz\ and \spec\ methods, ordered by shape catalog, \pz\ used in tomographic binning, and redshift range. Results for the fiducial sample are displayed in Figure~\ref{fig:fiducial_cosmos}. Results for different \pz\ algorithms run on the same data are shown in Table~\ref{tab:desy1_full}. The \wz\ measurements do not extend beyond $z=0.9$, and so a \wz\ calibration of the last tomographic bin is not performed. Calibrations are consistent between the two methods.
}
\label{tab:desy1}
\end{table*}

\section{Conclusion}
\label{sec:conclusions}

We have presented the \wz\ measurements of the DES Y1 source galaxy sample. Using these measurements, we calibrated the mean redshifts of binned source galaxies to the $\sim \pm 0.020$ level. When combined with \spec\ calibrations, we measure the mean redshifts to $\sim \pm 0.015$ \citep{photoz}. These uncertainties are a significant but subdominant contribution to the error budget of the DES Y1 cosmological parameter measurements.
Analysis of the DES three-year (Y3) data is underway.
At almost four times the area, greater depth, and more sophisticated calibration techniques, we expect a larger than twofold decrease in our statistical uncertainties over our current Y1 efforts. Unless our redshift distribution measurements improve accordingly, uncertainties in them may become a major source of uncertainty in DES Y3 and other future imaging surveys. It may no longer be sufficient to calibrate the mean redshifts, and we may need to consider the width and skewness, if not more sophisticated parameterisations of the redshift distribution.
Clustering redshifts may need improvement as well. \redmagic\ does not extend to high enough redshift, and so additional reference sources may need to be incorporated. The effectiveness of \wz\ for calibrating redshift distributions is severely diminished by both the evolution of bias within a tomographic bin and also the width of that bin. Both can be mitigated by using \wz\ and \pz\ properties to create galaxy samples with tight tomographic bins.
Clustering redshifts have an already significant part to play in mitigating redshift systematics. We expect clustering redshifts, and particularly their calibration of redshift distributions, will remain a critical area of research in the coming years.

\section*{Acknowledgements}

\input{acknowledgments.tex}




\bibliographystyle{mnras}
\bibliography{xcorr,des_y1kp_short}



\appendix


\section*{Affiliations}
\input{affiliations.tex}

\bsp	
\label{lastpage}
\end{document}

%% file: authors.tex
\author[DES Collaboration]{
\parbox{\textwidth}{
\Large
C.~Davis$^{1}$,
M.~Gatti$^{2}$,
P.~Vielzeuf$^{2}$,
R.~Cawthon$^{3}$,
E.~Rozo$^{4}$,
A.~Alarcon$^{5}$,
G.~M.~Bernstein$^{6}$,
C.~Bonnett$^{2}$,
A.~Carnero~Rosell$^{7,8}$,
F.~J.~Castander$^{5}$,
C.~Chang$^{3}$,
L.~N.~da Costa$^{7,8}$,
T.~M.~Davis$^{9,10}$,
J.~De~Vicente$^{11}$,
J.~DeRose$^{12,1}$,
A.~Drlica-Wagner$^{13}$,
J.~Elvin-Poole$^{14}$,
E.~Gaztanaga$^{5}$,
D.~Gruen$^{1,15}$,
J.~Gschwend$^{7,8}$,
W.~G.~Hartley$^{16,17}$,
B.~Hoyle$^{18}$,
H.~Lin$^{13}$,
M.~A.~G.~Maia$^{7,8}$,
R.~Miquel$^{19,2}$,
R.~L.~C.~Ogando$^{7,8}$,
M.~M.~Rau$^{18}$,
A.~Roodman$^{1,15}$,
E.~S.~Rykoff$^{1,15}$,
I.~Sevilla-Noarbe$^{11}$,
M.~A.~Troxel$^{20,21}$,
R.~H.~Wechsler$^{12,1,15}$,
T.~M.~C.~Abbott$^{22}$,
F.~B.~Abdalla$^{16,23}$,
S.~Allam$^{13}$,
J.~Annis$^{13}$,
K.~Bechtol$^{24}$,
A.~Benoit-L{\'e}vy$^{25,16,26}$,
D.~Brooks$^{16}$,
E.~Buckley-Geer$^{13}$,
D.~L.~Burke$^{1,15}$,
M.~Carrasco~Kind$^{27,28}$,
J.~Carretero$^{2}$,
M.~Crocce$^{5}$,
C.~E.~Cunha$^{1}$,
S.~Desai$^{29}$,
H.~T.~Diehl$^{13}$,
P.~Doel$^{16}$,
T.~F.~Eifler$^{30,31}$,
B.~Flaugher$^{13}$,
J.~Frieman$^{13,3}$,
J.~Garc\'ia-Bellido$^{32}$,
D.~W.~Gerdes$^{33,34}$,
R.~A.~Gruendl$^{27,28}$,
G.~Gutierrez$^{13}$,
K.~Honscheid$^{20,21}$,
D.~J.~James$^{35}$,
T.~Jeltema$^{36}$,
E.~Krause$^{1}$,
R.~Kron$^{13,3}$,
K.~Kuehn$^{37}$,
N.~Kuropatkin$^{13}$,
O.~Lahav$^{16}$,
M.~Lima$^{38,7}$,
M.~March$^{6}$,
J.~L.~Marshall$^{39}$,
F.~Menanteau$^{27,28}$,
R.~C.~Nichol$^{40}$,
B.~Nord$^{13}$,
A.~A.~Plazas$^{31}$,
E.~Sanchez$^{11}$,
V.~Scarpine$^{13}$,
R.~Schindler$^{15}$,
M.~Smith$^{41}$,
M.~Soares-Santos$^{13}$,
F.~Sobreira$^{42,7}$,
E.~Suchyta$^{43}$,
M.~E.~C.~Swanson$^{28}$,
G.~Tarle$^{34}$,
D.~Thomas$^{40}$,
D.~L.~Tucker$^{13}$,
V.~Vikram$^{44}$,
A.~R.~Walker$^{22}$,
J.~Zuntz$^{45}$
\begin{center} (DES Collaboration) \end{center}
}
\vspace{0.4cm}
\\
}

%% file: acknowledgments.tex
CPD is supported by the Northern California Chapter of the ARCS Foundation, as well as by the U.S. Department of Energy under contract number DE-AC02-76-SF00515.
ER acknowledges support by the DOE Early Career Program, DOE grant DE-SC0015975, and the Sloan Foundation, grant FG-2016-6443.

Funding for the DES Projects has been provided by the U.S. Department of Energy, the U.S. National Science Foundation, the Ministry of Science and Education of Spain, 
the Science and Technology Facilities Council of the United Kingdom, the Higher Education Funding Council for England, the National Center for Supercomputing 
Applications at the University of Illinois at Urbana-Champaign, the Kavli Institute of Cosmological Physics at the University of Chicago, 
the Center for Cosmology and Astro-Particle Physics at the Ohio State University,
the Mitchell Institute for Fundamental Physics and Astronomy at Texas A\&M University, Financiadora de Estudos e Projetos, 
Funda{\c c}{\~a}o Carlos Chagas Filho de Amparo {\`a} Pesquisa do Estado do Rio de Janeiro, Conselho Nacional de Desenvolvimento Cient{\'i}fico e Tecnol{\'o}gico and 
the Minist{\'e}rio da Ci{\^e}ncia, Tecnologia e Inova{\c c}{\~a}o, the Deutsche Forschungsgemeinschaft and the Collaborating Institutions in the Dark Energy Survey. 

The Collaborating Institutions are Argonne National Laboratory, the University of California at Santa Cruz, the University of Cambridge, Centro de Investigaciones Energ{\'e}ticas, 
Medioambientales y Tecnol{\'o}gicas-Madrid, the University of Chicago, University College London, the DES-Brazil Consortium, the University of Edinburgh, 
the Eidgen{\"o}ssische Technische Hochschule (ETH) Z{\"u}rich, 
Fermi National Accelerator Laboratory, the University of Illinois at Urbana-Champaign, the Institut de Ci{\`e}ncies de l'Espai (IEEC/CSIC), 
the Institut de F{\'i}sica d'Altes Energies, Lawrence Berkeley National Laboratory, the Ludwig-Maximilians Universit{\"a}t M{\"u}nchen and the associated Excellence Cluster Universe, 
the University of Michigan, the National Optical Astronomy Observatory, the University of Nottingham, The Ohio State University, the University of Pennsylvania, the University of Portsmouth, 
SLAC National Accelerator Laboratory, Stanford University, the University of Sussex, Texas A\&M University, and the OzDES Membership Consortium.

Based in part on observations at Cerro Tololo Inter-American Observatory, National Optical Astronomy Observatory, which is operated by the Association of 
Universities for Research in Astronomy (AURA) under a cooperative agreement with the National Science Foundation.

The DES data management system is supported by the National Science Foundation under Grant Numbers AST-1138766 and AST-1536171.
The DES participants from Spanish institutions are partially supported by MINECO under grants AYA2015-71825, ESP2015-88861, FPA2015-68048, SEV-2012-0234, SEV-2016-0597, and MDM-2015-0509, 
some of which include ERDF funds from the European Union. IFAE is partially funded by the CERCA program of the Generalitat de Catalunya.
Research leading to these results has received funding from the European Research
Council under the European Union's Seventh Framework Program (FP7/2007-2013) including ERC grant agreements 240672, 291329, and 306478.
We  acknowledge support from the Australian Research Council Centre of Excellence for All-sky Astrophysics (CAASTRO), through project number CE110001020.

This manuscript has been authored by Fermi Research Alliance, LLC under Contract No. DE-AC02-07CH11359 with the U.S. Department of Energy, Office of Science, Office of High Energy Physics. The United States Government retains and the publisher, by accepting the article for publication, acknowledges that the United States Government retains a non-exclusive, paid-up, irrevocable, world-wide license to publish or reproduce the published form of this manuscript, or allow others to do so, for United States Government purposes.

%% file: affiliations.tex
$^{1}$ Kavli Institute for Particle Astrophysics \& Cosmology, P. O. Box 2450, Stanford University, Stanford, CA 94305, USA\\
$^{2}$ Institut de F\'{\i}sica d'Altes Energies (IFAE), The Barcelona Institute of Science and Technology, Campus UAB, 08193 Bellaterra (Barcelona) Spain\\
$^{3}$ Kavli Institute for Cosmological Physics, University of Chicago, Chicago, IL 60637, USA\\
$^{4}$ Department of Physics, University of Arizona, Tucson, AZ 85721, USA\\
$^{5}$ Institute of Space Sciences, IEEC-CSIC, Campus UAB, Carrer de Can Magrans, s/n,  08193 Barcelona, Spain\\
$^{6}$ Department of Physics and Astronomy, University of Pennsylvania, Philadelphia, PA 19104, USA\\
$^{7}$ Laborat\'orio Interinstitucional de e-Astronomia - LIneA, Rua Gal. Jos\'e Cristino 77, Rio de Janeiro, RJ - 20921-400, Brazil\\
$^{8}$ Observat\'orio Nacional, Rua Gal. Jos\'e Cristino 77, Rio de Janeiro, RJ - 20921-400, Brazil\\
$^{9}$ ARC Centre of Excellence for All-sky Astrophysics (CAASTRO)\\
$^{10}$ School of Mathematics and Physics, University of Queensland,  Brisbane, QLD 4072, Australia\\
$^{11}$ Centro de Investigaciones Energ\'eticas, Medioambientales y Tecnol\'ogicas (CIEMAT), Madrid, Spain\\
$^{12}$ Department of Physics, Stanford University, 382 Via Pueblo Mall, Stanford, CA 94305, USA\\
$^{13}$ Fermi National Accelerator Laboratory, P. O. Box 500, Batavia, IL 60510, USA\\
$^{14}$ Jodrell Bank Center for Astrophysics, School of Physics and Astronomy, University of Manchester, Oxford Road, Manchester, M13 9PL, UK\\
$^{15}$ SLAC National Accelerator Laboratory, Menlo Park, CA 94025, USA\\
$^{16}$ Department of Physics \& Astronomy, University College London, Gower Street, London, WC1E 6BT, UK\\
$^{17}$ Department of Physics, ETH Zurich, Wolfgang-Pauli-Strasse 16, CH-8093 Zurich, Switzerland\\
$^{18}$ Universit\"ats-Sternwarte, Fakult\"at f\"ur Physik, Ludwig-Maximilians Universit\"at M\"unchen, Scheinerstr. 1, 81679 M\"unchen, Germany\\
$^{19}$ Instituci\'o Catalana de Recerca i Estudis Avan\c{c}ats, E-08010 Barcelona, Spain\\
$^{20}$ Center for Cosmology and Astro-Particle Physics, The Ohio State University, Columbus, OH 43210, USA\\
$^{21}$ Department of Physics, The Ohio State University, Columbus, OH 43210, USA\\
$^{22}$ Cerro Tololo Inter-American Observatory, National Optical Astronomy Observatory, Casilla 603, La Serena, Chile\\
$^{23}$ Department of Physics and Electronics, Rhodes University, PO Box 94, Grahamstown, 6140, South Africa\\
$^{24}$ LSST, 933 North Cherry Avenue, Tucson, AZ 85721, USA\\
$^{25}$ CNRS, UMR 7095, Institut d'Astrophysique de Paris, F-75014, Paris, France\\
$^{26}$ Sorbonne Universit\'es, UPMC Univ Paris 06, UMR 7095, Institut d'Astrophysique de Paris, F-75014, Paris, France\\
$^{27}$ Department of Astronomy, University of Illinois, 1002 W. Green Street, Urbana, IL 61801, USA\\
$^{28}$ National Center for Supercomputing Applications, 1205 West Clark St., Urbana, IL 61801, USA\\
$^{29}$ Department of Physics, IIT Hyderabad, Kandi, Telangana 502285, India\\
$^{30}$ Department of Physics, California Institute of Technology, Pasadena, CA 91125, USA\\
$^{31}$ Jet Propulsion Laboratory, California Institute of Technology, 4800 Oak Grove Dr., Pasadena, CA 91109, USA\\
$^{32}$ Instituto de Fisica Teorica UAM/CSIC, Universidad Autonoma de Madrid, 28049 Madrid, Spain\\
$^{33}$ Department of Astronomy, University of Michigan, Ann Arbor, MI 48109, USA\\
$^{34}$ Department of Physics, University of Michigan, Ann Arbor, MI 48109, USA\\
$^{35}$ Astronomy Department, University of Washington, Box 351580, Seattle, WA 98195, USA\\
$^{36}$ Santa Cruz Institute for Particle Physics, Santa Cruz, CA 95064, USA\\
$^{37}$ Australian Astronomical Observatory, North Ryde, NSW 2113, Australia\\
$^{38}$ Departamento de F\'isica Matem\'atica, Instituto de F\'isica, Universidade de S\~ao Paulo, CP 66318, S\~ao Paulo, SP, 05314-970, Brazil\\
$^{39}$ George P. and Cynthia Woods Mitchell Institute for Fundamental Physics and Astronomy, and Department of Physics and Astronomy, Texas A\&M University, College Station, TX 77843,  USA\\
$^{40}$ Institute of Cosmology \& Gravitation, University of Portsmouth, Portsmouth, PO1 3FX, UK\\
$^{41}$ School of Physics and Astronomy, University of Southampton,  Southampton, SO17 1BJ, UK\\
$^{42}$ Instituto de F\'isica Gleb Wataghin, Universidade Estadual de Campinas, 13083-859, Campinas, SP, Brazil\\
$^{43}$ Computer Science and Mathematics Division, Oak Ridge National Laboratory, Oak Ridge, TN 37831\\
$^{44}$ Argonne National Laboratory, 9700 South Cass Avenue, Lemont, IL 60439, USA\\
$^{45}$ Institute for Astronomy, University of Edinburgh, Edinburgh EH9 3HJ, UK\\